\documentclass[aps,preprint]{revtex4-1}
\usepackage{amsmath}
\usepackage{amsfonts}
\usepackage{amssymb}
\usepackage{graphicx}%
\providecommand{\U}[1]{\protect \rule{.1in}{.1in}}
\begin{document}

%

\title{Anharmonicity-Induced Criticality of Collective Excitation in a Trapped
Bose-Einstein Condensate}
\author{Qun Wang$^{\ast}$ and Bo Xiong$^{\ast},^{\dagger},^{\ddagger}$}

\address{$^\ast$Department of Physics, Nanchang University,\\
330031 Nanchang, China}
\address{$^\dagger$Skolkovo Institute of Science and Technology,\\
Novaya Street 100, Skolkovo 143025, Russian Federation}
\address{$^\ddagger$stevenxiongbo@gmail.com}


\begin{abstract}
We investigate the low energy excitations of a dilute atomic Bose gas confined
in a anharmonic trap interacting with repulsive forces. The dispersion law of
both surface and compression modes are derived and analyzed for large numbers
of atoms in the trap, which show two branches of excitation and appear a two
critical value. For a upper limit, BEC can be unstable with respect to some
specific collective excitation, while for the lower limit, the frequency of
collective excitation under anharmonic influence can be effectively lower than
that without anharmonicity. Our work reveals the key role played by the
anharmonicity and interatomic forces which introduce a rich structure in the
dynamic behavior of these new many-body systems.

\end{abstract}


\maketitle

\section{Introduction}

Bose-Einstein Condensates (BEC) is a macroscopic quantum phenomenon that
interactions between atoms strongly affect their properties such as
superfluid, superconductivity and so on. Elementary excitations play a crucial
role in the understanding of many-body quantum systems, which gain insights
into these physical properties of trapped BEC \cite{1,2,3,4,5,6,6-1}. In 1947,
Bogoliubov \cite{7} put forward his famous theory for collective elementary
excitations of the dilute homogeneous Bose gas \cite{8,9} which is
corresponding to the linear limit of the Gross-Pitaevskii equation (GPE)
\cite{10} with order parameters and is well suitable for study of the effects
of weak interactions on the formation of BEC in dilute homogeneous or
inhomogeneous Bose gases at low temperature \cite{3,8,11}.

The observation of BEC in a remarkable series of experiments, such as on
vapors of rubidium by Eric Cornell, Carl Wieman \cite{14}, sodium by Wolfgang
Ketterle \cite{15}, lithium by Randall Hulet \cite{15-1}, and so on \cite{16}
has already led to advances in our understanding of the weakly interacting
Bose gas. Since then, collective elementary excitations have attracted much
interest in both experimental and theoretical methods \cite{6-1,11}.

Stringari has investigated low energy elementary excitations of dilute Bose
gas confined in a harmonic trap employing the hydrodynamic approximation and
the sum rule approach \cite{8,11,17}. For isotropic harmonic oscillator trap,
the calculation results of quadrupole mode $(n=0,\ell=2,m=0)$ and monopole
mode $(n=1,\ell=0,m=0)$ are $\omega_{Q}=\sqrt{2}\omega_{0}$ and $\omega
_{M}=\sqrt{5}\omega_{0}$, which match the findings of experiment~\cite{18}. In
most practical situations of experiments, the confining trap is usually
anisotropic \cite{18,19,20,21,22,24,25,28}. For the anisotropic harmonic trap
with axial symmetry, the calculation results of quadrupole mode are
$\omega=\sqrt{2}\omega_{\bot}$ or $\omega=\sqrt{\omega_{\bot}^{2}+\omega
_{z}^{2}}$ for the $m=\pm2$ and $m=\pm1$, respectively. In particular,
quadrupole mode $(m=0)$ has a coupling dynamics with monopole mode
$(n=1,\ell=0)$ with frequency $\omega=1.797\omega_{\bot}$ for decoupled modes,
for disk-type anisotropy $\omega_{z}/\omega_{\bot}=\sqrt{8}$ which is shown
that the theoretical and numerical results \cite{18,23} are excellent agree
with the experiment results~\cite{24}. On the other hand, for cigar-type
anisotropy $\omega_{z}/\omega_{\bot}\ll1$, the frequency of decoupled modes is
$\omega=\sqrt{5/2}\omega_{z}$ or $\omega=2\omega_{\bot}$ which are agree with
the experiment results \cite{25} excellently. Recently, experimentally
realized the study and control of elementary excitations confined in an
isotropic harmonic potential~\cite{27-1,27} has successfully proven the highly
suppressed damping for the monopole mode. The observed frequencies of two
elementary collective modes are $\omega_{Q}=1.435\omega_{0}$ and $\omega
_{M}=2.283\omega_{0}$ for quadrupole and monopole mode respectively in
\cite{27}, which are consistent with theoretical results \cite{18}.

In any practical situation, the trapping potential is not accurately harmonic.
It is important to notice that the anharmonicity can significantly cause
center-of-mass and relative motion coupling \cite{31,32}, where
anharmonicity-induced resonances have been investigated and confirmed in
ultracold few-body bosonic systems confined in anharmonic trap experimentally
\cite{30,30-1}. Moreover, these anharmonicity-induced inelastic
confinement-induced resonances can lead to coherent molecule formation,
losses, and heating in ultracold atomic gases, which can be readily reached
upon the state-of-the-art experimental tools and techniques \cite{30,30-1}.

In this paper, we investigate collective excitation spectrum of BEC in an 3D
harmonic trap under the anharmonic perturbation. We find that, the dynamics of
elementary excitations deeply affected by anharmonicity $\tilde{\eta}$, which
has a two critical value, upper one of which represents instability collective
excitation, and lower one of which which represents critical value at which
the frequency of collective excitation with influence by anharmonicity
$\tilde{\eta}$ is lower than that without anharmonicity. Moreover, the
dispersion law is now separated into two branches, which is quite different
from the harmonic trapping case where, for the surface modes, the frequency of
lower branch can be smaller than that without anharmonicity, for the
compression modes, both branches can be smaller than those without
anharmonicity surprisingly.

This article is organized as follows. In Sec. \ref{D}, we introduce the
collective excitations of a Bose-condensed gas trapped in isotropic harmonic
oscillator potential. With basis of Sec. \ref{D}, in Sec. \ref{E}, we study
the collective excitations spectrum in anisotropic harmonic oscillator
potential with or without axial symmetry. In Sec. \ref{F}, anharmonic
perturbation has been considered together with harmonic trapping potential.
Discussion and conclusions are stated in Sec. \ref{G}.

\section{Isotropic Harmonic potential}

\label{D}

For dilute atomic Bose gas at zero temperature, the macroscopic wave function
$\Phi \left(  \mathbf{r},t\right)  $ satisfies the time-dependent GPE \cite{10}
in the following:
\begin{equation}
i\hbar \frac{\partial}{\partial t}\Phi \left(  \mathbf{r},t\right)  =\left(
-\frac{\hbar^{2}\nabla^{2}}{2m}+V_{\text{ext}}\left(  \mathbf{r}\right)
+g\mid \Phi \left(  \mathbf{r},t\right)  \mid^{2}\right)  \Phi \left(
\mathbf{r},t\right)  , \label{1}%
\end{equation}
here $g=4\pi \hbar^{2}a/m$ is the interaction coupling constant, $a$ is the
s-wave scattering length and $m$ is the atom mass, the $V_{ext}\left(
\mathbf{r}\right)  $ is the external confining potential. We can obtain the
collective excitations of the Bose-condensed gas analytically when the
interaction is large enough to make the kinetic energy pressure negligible
compared to the external and interparticle interaction terms.

When the Bose-Condensed gas in a ground state, i.e., $i\hbar \frac{\partial
}{\partial t}\Phi \left(  \mathbf{r},t\right)  =\mu \Phi \left(  \mathbf{r}%
,t\right)  $ where $\mu$ is chemical potential, the Eq. (\ref{1}) can be
rewritten as:
\begin{equation}
\left(  -\frac{\hbar^{2}\nabla^{2}}{2m}+V_{\text{ext}}\left(  \mathbf{r}%
\right)  +g\mid \Phi \left(  \mathbf{r},t\right)  \mid^{2}\right)  \Phi \left(
\mathbf{r},t\right)  =\mu \Phi \left(  \mathbf{r},t\right)  . \label{2}%
\end{equation}

In order to discuss the dispersion spectrum of the elementary excitations,
explicit equations for the density $\rho \left(  \mathbf{r},t\right)  =\mid
\Phi \left(  \mathbf{r},t\right)  \mid^{2}$ and the velocity field $V\left(
\mathbf{r},t\right)  =\hbar \left[  \Phi^{\ast}\left(  \mathbf{r},t\right)
\nabla \Phi \left(  \mathbf{r},t\right)  -\Phi \left(  \mathbf{r},t\right)
\nabla \Phi^{\ast}\left(  \mathbf{r},t\right)  \right]  /2mi\rho \left(
\mathbf{r},t\right)  $ have been derived as follows. By inserting $\Phi \left(
\mathbf{r},t\right)  =\sqrt{\rho}e^{i\left(  \theta \left(  \mathbf{r}%
,t\right)  -\mu t/\hbar \right)  }$ into Eq. (\ref{1}), we can get following
Eqs.:%
\begin{equation}
\frac{\partial}{\partial t}\rho+\nabla \cdot \left(  \mathbf{V}\rho \right)  =0,
\label{4}%
\end{equation}%
\begin{equation}
\frac{\partial}{\partial t}\mathbf{V}+\nabla \left(  \delta \mu+\frac{1}%
{2}m\mathbf{V}^{2}\right)  =0, \label{5}%
\end{equation}
here velocity field is%
\begin{equation}
\mathbf{V}=\frac{\hbar}{m}\nabla \theta, \label{3}%
\end{equation}
and%
\begin{equation}
\delta \mu=V_{\text{ext}}\left(  \mathbf{r}\right)  +\frac{4\pi \hbar^{2}a}%
{m}\rho-\frac{\hbar^{2}}{2m\sqrt{\rho}}\nabla^{2}\sqrt{\rho}-\mu \label{6}%
\end{equation}
is the change of the chemical potential with respect to its ground state value
\cite{8}. The Eq. (\ref{3}) and Eq. (\ref{5}) has reflected irrotational
nature of the superfluid motion \cite{11}. By setting $\mathbf{V}=0$ and
$\delta \mu=0$, superfluid density $\rho_{0}$ relative to the ground state can
be obtained as \cite{11}:
\begin{equation}
V_{\text{ext}}\left(  \mathbf{r}\right)  +\frac{4\pi \hbar^{2}a}{m}\rho
_{0}-\frac{\hbar^{2}}{2m\sqrt{\rho_{0}}}\nabla^{2}\sqrt{\rho_{0}}-\mu=0,
\label{7}%
\end{equation}
which coincides with GPE for the macroscopic wave function $\Phi_{0}\left(
\mathbf{r},t\right)  =\sqrt{\rho_{0}}$ of the ground state. Chemical potential
can be fixed by requiring the density $\rho_{0}$ normalized to some value
which can be adjusted to some extent experimentally. When the number of the
Bose atoms is large enough, interaction energy between atoms is much larger
than kinetic energy, and the density profile $\rho_{0}\left(  \mathbf{r}%
\right)  $ become smooth \cite{8}, so that the kinetic energy pressure term
$\left(  \hbar^{2}/2m\sqrt{\rho}\right)  \nabla^{2}\sqrt{\rho}$ can be
neglected with respect to the interaction terms of Eq. (\ref{7}). and then,
the well known Thomas-Fermi approximation for the ground state density can be
obtained as \cite{8}:
\begin{equation}
\rho_{0}\left(  \mathbf{r}\right)  =\frac{m}{4\pi \hbar^{2}a}\left[
\mu-V_{ext}\left(  \mathbf{r}\right)  \right]  , \label{8}%
\end{equation}
if $\mu \geq V_{\text{ext}}\left(  \mathbf{r}\right)  $ and is equal to zero
elsewhere. In this case, Eq.(\ref{6}) and Eq. (\ref{7}) can be combined
together as
\[
\delta \mu=\frac{4\pi \hbar^{2}a}{m}\left[  \rho \left(  \mathbf{r},t\right)
-\rho_{0}\left(  \mathbf{r}\right)  \right]  ,
\]
for the change of the chemical potential \cite{8}. In the case of isotropic
harmonic trapping potential $V_{\text{ext}}\left(  \mathbf{r}\right)
=m\omega_{0}^{2}r^{2}/2$, the equations of motion (\ref{4}) and (\ref{5}),
after linearization around ground state density $\rho_{0}$, can be written in
the following form:
\begin{equation}
-\omega^{2}\delta \rho e^{-i\omega t}=\frac{\partial^{2}}{\partial t^{2}}%
\rho \left(  \mathbf{r},t\right)  =\nabla \cdot \left(  \rho_{0}\nabla \delta \rho
e^{-i\omega t}\right)  +T_{1}+T_{2}, \label{10}%
\end{equation}
here, total density $\rho \left(  \mathbf{r},t\right)  =\rho_{0}\left(
\mathbf{r}\right)  +\delta \rho e^{-i\omega t}$ and high order smaller term
$T_{1}=\nabla \cdot \left(  \delta \rho e^{-i\omega t}\frac{g}{m}\nabla \delta \rho
e^{-i\omega t}\right)  $ and velocity-dependent term $T_{2}=\left[  \frac
{1}{2}\nabla \cdot \left(  \rho_{0}\nabla \mathbf{V}^{2}\right)  +\nabla
\cdot \left(  i\omega \delta \rho e^{-i\omega t}\mathbf{V}\right)  +\frac{1}%
{2}\nabla \cdot \left(  \delta \rho e^{-i\omega t}\nabla \mathbf{V}^{2}\right)
\right]  $ which can be neglected when the values of dimensionless parameter
$Na/a_{HO}$ is sufficiently large (\cite{8,11}), thus, the Eq.(\ref{10}) have
a simple form with harmonic trapping potential:
\begin{equation}
\omega^{2}\delta \rho=-\nabla \cdot \left(  \rho_{0}\nabla \delta \rho \right)
=-\frac{1}{2}\omega_{0}^{2}\nabla \cdot \left[  \left(  R^{2}-r^{2}\right)
\cdot \nabla \delta \rho \right]  , \label{11}%
\end{equation}
where $R^{2}=2\mu/m\omega_{0}^{2}$ with dimension length$^{2}$. The
hydrodynamic Eq. (\ref{11}) are defined in the interval $0\leq r\leq R$ and
have solutions as following form:
\begin{equation}
\delta \rho \left(  \mathbf{r}\right)  =P_{\ell}^{2n}\left(  \frac{r}{R}\right)
r^{\ell}Y_{\ell m}\left(  \theta,\phi \right)  , \label{PolyEP}%
\end{equation}
among them, $P_{\ell}^{\left(  2n\right)  }\left(  t\right)  =\sum_{k=0}%
^{n}\alpha_{2k}t^{2k}$ are polynomials of degree $2n$ and $\alpha_{0}=1$, only
containing the even powers of $t$, and satisfying the orthogonality condition
$\int_{0}^{1}P_{\ell}^{2n}\left(  t\right)  P_{\ell}^{2n^{\prime}}\left(
t\right)  t^{2\ell+2}dt=0$ if $n\neq n^{\prime}$ \cite{8}. Here $2n$ is the
number of radial nodes, $\ell,m$ is the quantum number of orbital angular
momentum and that z-component of the excitation, respectively. In the
following, we will give derivation processes for the excitation spectrum based
on the (\ref{PolyEP}) under the harmonic trapping potential.

By defining $t=r/R$ and polynomials $T_{\ell}^{2n}\left(  t\right)  =P_{\ell
}^{2n}\left(  t=\frac{r}{R}\right)  r^{\ell}$, then $T_{\ell}^{2n}\left(
t\right)  $ has following form
\begin{equation}
T_{\ell}^{2n}\left(  t\right)  =R^{\ell}t^{\ell}\left(  1+\alpha_{2}%
t^{2}+\cdots+\alpha_{2n}t^{2n}\right)  =\sum_{k=o}^{n}R^{\ell}\alpha
_{2k}t^{2k+\ell}, \label{13}%
\end{equation}
and orthogonality condition for $T_{\ell}^{2n}\left(  t\right)  $%
\begin{equation}
\int_{0}^{1}T_{\ell}^{2n}\left(  t\right)  T_{\ell}^{2n^{\prime}}\left(
t\right)  t^{2}dt=0,\text{ \ }\left(  n\neq n^{\prime}\right)  , \label{14}%
\end{equation}
by using the Eq. (\ref{11})--Eq. (\ref{13}), we can obtain the following Eq.:
\begin{equation}
Y_{\ell m}\sum_{k=0}^{n}\left(  \mathbf{B+C}\right)  =0, \label{15}%
\end{equation}
here%
\begin{gather}
\mathbf{B}=R^{\ell}\sum_{k=0}^{n}\left[  \frac{2\omega^{2}}{\omega_{0}%
}-2\left(  2k^{2}+3k+2k\ell+2\ell \right)  \right]  \alpha_{2k}t^{2k+\ell
}\nonumber \\
=b_{0}T_{\ell}^{0}+b_{1}T_{\ell}^{2}+\cdots+b_{k}T_{\ell}^{2k}+\cdots
+b_{n}T_{\ell}^{2n}, \label{16}%
\end{gather}
and
\begin{gather}
\mathbf{C}=R^{\ell}\sum_{k=0}^{n}\left[  2\left(  2k^{2}+k+2k\ell \right)
\alpha_{2k}t^{2k+\ell-2}\right] \nonumber \\
=c_{0}T_{\ell}^{0}+c_{1}T_{\ell}^{2}+\cdots+c_{k}T_{\ell}^{2k}+\cdots
+c_{n-1}T_{\ell}^{2n-2}. \label{17}%
\end{gather}
By equating the coefficients of same order of $T$ in Eq. (\ref{16}) and Eq.
(\ref{17}), we can get following equations:%
\begin{gather}
\left(  b_{0}+b_{1}+\cdots+b_{k}+\cdots+b_{n}\right)  \alpha_{0}=\left[
\frac{2\omega^{2}}{\omega_{0}^{2}}-2\ell \right]  \alpha_{0},\nonumber \\
\left(  0+b_{1}+\cdots+b_{k}+\cdots+b_{n}\right)  \alpha_{2}=\left[
\frac{2\omega^{2}}{\omega_{0}^{2}}-2\left(  2\times1^{2}+2\times1\ell
+3\times1+\ell \right)  \right]  \alpha_{2},\nonumber \\
\vdots \nonumber \\
\left(  0+0+\cdots+b_{k}+b_{k+1}+\cdots+b_{n}\right)  \alpha_{2k}=\left[
\frac{2\omega^{2}}{\omega_{0}^{2}}-2\left(  2k^{2}+2\times k\ell+3\times
k+\ell \right)  \right]  \alpha_{2k},\nonumber \\
\vdots \nonumber \\
\left(  0+0+\cdots+b_{n}\right)  \alpha_{2n}=\left[  \frac{2\omega^{2}}%
{\omega_{0}^{2}}-2\left(  2n^{2}+2n\ell+3n+\ell \right)  \right]  \alpha_{2n},
\label{18}%
\end{gather}
and%
\begin{gather}
\left(  c_{0}+c_{1}+\cdots+c_{k}+\cdots+c_{n-1}\right)  \alpha_{0}%
=2\times1\left(  2\times1+2\ell+1\right)  \alpha_{2}\nonumber \\
\left(  0+c_{1}+\cdots+c_{k}+\cdots+c_{n-1}\right)  \alpha_{2}=2\times2\left(
2\times2+2\ell+1\right)  \alpha_{4}\nonumber \\
\vdots \nonumber \\
\left(  0+0+\cdots+c_{k}+c_{k+1}+\cdots+c_{n-1}\right)  \alpha_{2k}=2\left(
k+1\right)  \left(  2k+2\ell+3\right)  \alpha_{2k+2}\nonumber \\
\vdots \nonumber \\
\left(  0+0+\cdots+c_{n-1}\right)  \alpha_{2n-2}=2n\left(  2n+2\ell+1\right)
\alpha_{2n}. \label{19}%
\end{gather}
By using orthogonality condition (\ref{14}), we can immediately obtain the
dispersion law of the normal modes as (\cite{8,11}):
\begin{equation}
\omega \left(  n,\ell \right)  =\omega_{0}\left(  2n^{2}+2n\ell+3n+\ell \right)
^{\frac{1}{2}}, \label{21}%
\end{equation}
and the recurrence relation for the coefficients $\alpha_{2k}$ can be derived
as:
\begin{gather}
\alpha_{2k+2}=\frac{\left(  c_{k}+c_{k+1}+\cdots+c_{n-1}\right)  }{2\left(
k+1\right)  \left(  2\ell+2k+3\right)  }\alpha_{2k}\nonumber \\
=-\frac{\left(  b_{k}+b_{k+1}+\cdots+b_{n-1}\right)  }{2\left(  k+1\right)
\left(  2\ell+2k+3\right)  }\alpha_{2k}\nonumber \\
=-\frac{\left(  n-k\right)  \left(  2\ell+2k+3+2n\right)  }{\left(
k+1\right)  \left(  2\ell+2k+3\right)  }\alpha_{2k}. \label{23}%
\end{gather}

\subsection{Discussion}

The dispersion law (\ref{21}) and recurrence relation (\ref{23}) of
coefficients $\alpha_{2k}$ are discussed in an isotropic harmonic oscillator
potential (\cite{8,11}). The lowest radial modes $\left(  n=0\right)  $ (also
called surface excitations) and its dispersion laws are given by (\ref{21}):
\begin{equation}
\omega \left(  n=0\right)  =\sqrt{\ell}\omega_{0}, \label{25}%
\end{equation}
the frequency of these modes lies systematically below the harmonic oscillator
result $\omega_{\text{HO}}\left(  n=0\right)  =\ell \omega_{0}$. It is very
interesting to notice that the dispersion law (\ref{21}) predicting quadrupole
frequency $\omega_{Q}(n=0,\ell=2,m=0)=\sqrt{2}\omega_{0}$, monopole frequency
$\omega_{M}(n=1,\ell=0,m=0)=\sqrt{5}\omega_{0}$ and radial-surface frequency
$\omega \left(  n=2,\ell=1,m=0\right)  =\sqrt{19}\omega_{0}$, are in good
agreement with recent experimental measurements $\omega_{Q}=1.435\omega_{0}$,
$\omega_{M}=2.283$ and $\omega \left(  n=2,\ell=1,m=0\right)  =4.361\omega_{0}$
in Ref. (\cite{27}) with isotropic harmonic trapping potential.

\section{Anisotropic Harmonic potential}

\label{E}

In the following, we will discuss effect of anisotropic harmonic potential on
the dispersion law of collective excitation, which are usually available with
magnetic traps \cite{8,19,20,21,22,28}, experimentally.

\subsection{Axial Symmetry}

First, we consider the case of a harmonic oscillator trap with axial symmetry
along the z axis: $V_{\text{ext}}=m\omega_{\perp}^{2}s^{2}/2+m\omega_{z}%
^{2}z^{2}$ and among them $s=\left(  x^{2}+y^{2}\right)  ^{\frac{1}{2}}$ is
the radial variable in the x-y plane. In this case the relevant hydrodynamic
Eq. (\ref{11}) can be rewritten as
\begin{equation}
\omega^{2}\delta \rho=-\frac{1}{2}\nabla \cdot \biggl \{ \left[  \omega_{\perp
}^{2}\left(  S^{2}-s^{2}\right)  +\omega_{z}\left(  Z^{2}-z^{2}\right)
\right]  \nabla \delta \rho \biggr \}, \label{27}%
\end{equation}
where $m\omega_{\perp}^{2}S^{2}/2+m\omega_{z}^{2}Z^{2}/2\equiv \mu$ (\cite{8}).

Since the trap is axisymmetric, the magnetic quantum number $m$ of angular
momentum is still a good quantum number. Very interestingly, however, the
dispersion law will depend on $m$ as shown in (\cite{8}). In some particular
cases $m=\pm \ell$ and $m=\pm \left(  \ell-1\right)  $, functions of the form
Eq. (\ref{PolyEP}) are still solutions of Eq. (\ref{27}) for $\delta \rho$,
resulting dispersion laws are derived as following: by inserting Eq.
(\ref{PolyEP}) into Eq. (\ref{27}), we obtain%
\begin{gather}
2R^{\ell}\sum_{k=0}^{n}[\omega^{2}Y_{\ell m}-\left(  2k^{2}+2k\ell
+3k+\ell \right)  \left(  \omega_{\bot}^{2}\sin^{2}\theta+\omega_{z}^{2}%
\cos^{2}\theta \right)  Y_{\ell m}\nonumber \\
-\left(  \omega_{\bot}^{2}-\omega_{z}^{2}\right)  \sin \theta \cos \theta
\frac{\partial}{\partial \theta}Y_{\ell m}]\alpha_{2k}t^{2k+\ell}\nonumber \\
+2R^{\ell}\sum_{k=0}^{n}\left(  \frac{S^{2}}{R^{2}}\omega_{\bot}^{2}%
+\frac{Z^{2}}{R^{2}}\omega_{z}^{2}\right)  Y_{\ell m}k\left(  2k+2\ell
+1\right)  \alpha_{2k}t^{2k+\ell-2}=0, \label{28}%
\end{gather}
when isotropic case $\omega_{\bot}=\omega_{z}$, the Eq.(\ref{28}) can be
reduced to Eq. (\ref{15}). By using the similar procedure as in Eq.(\ref{15}%
)--Eq.(\ref{20}), we can obtain the following Eq. determining the dispersion
law as
\begin{gather}
\omega^{2}Y_{\ell m}-\left(  2n^{2}+2n\ell+3n+\ell \right)  \left(
\omega_{\bot}^{2}\sin^{2}\theta+\omega_{z}^{2}\cos^{2}\theta \right)  Y_{\ell
m}\nonumber \\
-\left(  \omega_{\bot}^{2}-\omega_{z}^{2}\right)  \sin \theta \cos \theta
\frac{\partial}{\partial \theta}Y_{\ell m}=0, \label{29}%
\end{gather}
which can be solved by integrated out angular part in Eq.(\ref{29}) as:
\begin{gather}
\int_{0}^{2\pi}d\phi \int_{0}^{\pi}\sin \theta d\theta Y_{\ell m}^{\ast
}\biggl \{ \omega^{2}Y_{\ell m}-\left(  2n^{2}+2n\ell+3n+\ell \right)  \left(
\omega_{\bot}^{2}\sin^{2}\theta+\omega_{z}^{2}\cos^{2}\theta \right)  Y_{\ell
m}\nonumber \\
-\left(  \omega_{\bot}^{2}-\omega_{z}^{2}\right)  \sin \theta \cos \theta
\frac{\partial}{\partial \theta}Y_{\ell m}\biggr \}=0. \label{30}%
\end{gather}

First, we consider surface excitation $n=0$ with $m=\pm \ell$ or $m=\pm \left(
\ell-1\right)  $, and then, discuss general case. As we know the spherical
harmonic function is related with an associated Legendre polynomial as
\begin{equation}
Y_{\ell m}\left(  \theta,\phi \right)  =\left(  -1\right)  ^{m}N_{\ell}%
^{m}P_{\ell}^{m}\left(  \cos \theta \right)  e^{im\phi},\left(  -\ell \leq
m\leq \ell \right)  , \label{31}%
\end{equation}
here,%
\begin{equation}
N_{\ell}^{m}=\left(  -1\right)  ^{m}\left[  \frac{\left(  2\ell+1\right)
\cdot \left(  \ell-m\right)  !}{4\pi \cdot \left(  \ell+m\right)  !}\right]
^{\frac{1}{2}}, \label{33}%
\end{equation}
and the associated Legendre polynomials have property%
\begin{equation}
\int_{-1}^{1}P_{\ell}^{m}P_{\ell^{\prime}}^{m}dx=\frac{2\cdot \left(
\ell+m\right)  !}{\left(  2\ell+1\right)  \cdot \left(  \ell-m\right)  !},
\label{32}%
\end{equation}
and satisfy the following relationships:
\begin{equation}
\left(  1-x^{2}\right)  \frac{d}{dx}P_{\ell}^{m}\left(  x\right)  =\ell
xP_{\ell}^{m}\left(  x\right)  -\left(  \ell+m\right)  P_{\ell-1}^{m}\left(
x\right)  , \label{34}%
\end{equation}
and
\begin{equation}
\left(  1-x^{2}\right)  ^{\frac{1}{2}}P_{\ell}^{m}\left(  x\right)  =\left(
\frac{1}{2\ell+1}\right)  \left[  -P_{\ell+1}^{m+1}\left(  x\right)
+P_{\ell-1}^{m+1}\left(  x\right)  \right]  . \label{35}%
\end{equation}
Considering Eq.(30)--Eq.(35), the dispersion law of surface mode in a harmonic
oscillator trap with axial symmetry can be obtained from Eq. (\ref{29}) as
(\cite{8}):
\begin{equation}
\omega^{2}=F\left(  \ell,m\right)  \omega_{z}^{2}+G\left(  \ell,m\right)
\omega_{\bot}^{2}, \label{36}%
\end{equation}
here, $F\left(  \ell,m\right)  \equiv \left(  \ell+m\right)  \left(
\ell-m\right)  /\left(  2\ell-1\right)  $, $G\left(  \ell,m\right)
\equiv \left(  \ell^{2}+m^{2}-\ell \right)  /\left(  2\ell-1\right)  $, and
$m=\pm \ell$ or $m=\pm \left(  \ell-1\right)  $. Eq. (\ref{36}) can be reduced
to $\omega=\sqrt{\ell}\omega_{0}$ under the isotropic case $\omega_{\bot
}=\omega_{z}=\omega_{0}$. In addition, Eq. (\ref{36}) is well described the
dipole excitation $\left(  \ell=1\right)  $ whose frequencies coincide with
harmonic oscillator values $\omega_{D}\left(  m=\pm1\right)  =\omega_{\bot}$
and $\omega_{D}\left(  m=0\right)  =\omega_{z}$. But the quadrupole mode
($\ell=2$), Eq. (\ref{36}) can only describe the $\omega_{Q}\left(
m=\pm2\right)  =\sqrt{2}\omega_{\bot}$ and $\omega_{Q}\left(  m=\pm1\right)
=\left(  \omega_{\bot}^{2}+\omega_{z}^{2}\right)  ^{\frac{1}{2}}$ components.
These results have been derived in Ref. (\cite{8}) which gives excellent explanation.

In general case $n\neq0$, Eq. (\ref{29}) gives the dispersion law for the
collective excitation with radial node and surface pattern, by using the
similar way as $n=0$. The dispersion law we obtained as follows:
\begin{equation}
\omega^{2}=\ell \omega_{\bot}^{2}+E\left(  n,\ell \right)  \omega_{z}%
^{2}-F\left(  \ell,m\right)  \left(  \omega_{\bot}^{2}-\omega_{z}^{2}\right)
+E\left(  n,\ell \right)  H\left(  \ell,m\right)  \left(  \omega_{\bot}%
^{2}-\omega_{z}^{2}\right)  , \label{39}%
\end{equation}
here, $E\left(  n,\ell \right)  \equiv \left(  2n^{2}+2n\ell+3n\right)  $,
$H\left(  \ell,m\right)  \equiv \left(  \ell+m+2\right)  \left(  \ell
+m+1\right)  /\left(  \left(  2\ell+1\right)  \left(  2\ell+3\right)  \right)
+\left(  \ell-m\right)  \left(  \ell-m-1\right)  /\left(  \left(
2\ell+1\right)  \left(  2\ell-1\right)  \right)  $, and which can be reduced
to the same dispersion law as shown in Eq. (\ref{36}) when exciting lowest
radial mode $n=0$. Moreover, Eq. (\ref{39}) can be completely reduced to Eq.
(\ref{21}) under the isotropic case $\omega_{0}=\omega_{\bot}=\omega_{z}$.

\subsection{Coupling Modes with Axial Symmetry}

As mentioned in Ref. (\cite{8}), the monopole mode $\left(  n=1,\ell
=0,m=0\right)  $ and the quadrupole mode $\left(  n=0,\ell=2,m=0\right)  $ in
the axial symmetric harmonic trapping case are coupled together due to the
three-dimensional rotational symmetry broken, thus, functions of the form
$\delta \rho \left(  m=0\right)  $ is given by:
\begin{gather}
\delta \rho \left(  m=0\right)  =aP_{\ell}^{2}\left(  \frac{r}{R}\right)
+br^{2}Y_{2,0}\left(  \theta,\phi \right)  =a\left(  1+\alpha_{2}\frac{r^{2}%
}{R^{2}}\right)  +br^{2}\left(  12\cos^{2}\theta-4\right) \nonumber \\
=a+\left(  a\frac{\alpha_{2}}{R^{2}}-4b\right)  s^{2}+\left(  a\frac
{\alpha_{2}}{R^{2}}+8c\right)  z^{2}=a_{0}+b_{0}s^{2}+c_{0}z^{2}, \label{40}%
\end{gather}
where $a,b,c,a_{0},b_{0}$ and $c_{0}$ are constant coefficients which are
determined by the hydrodynamic Eq.(\ref{27}). Under the anisotropic harmonic
potential $\lambda=\omega_{z}/\omega_{\bot}$, by substituting Eq. (\ref{40})
into Eq. (\ref{27}), we can obtain the following equation as
\begin{gather}
\left[  \frac{\omega^{2}}{\omega_{\bot}^{2}}a_{0}+\left(  2b_{0}+c_{0}\right)
\left(  S^{2}-\lambda^{2}Z^{2}\right)  \right]  +\left[  \frac{\omega^{2}%
}{\omega_{\bot}^{2}}b_{0}-\left(  2b_{0}+c_{0}\right)  -2b_{0}\right]
s^{2}\nonumber \\
+\left[  \frac{\omega^{2}}{\omega_{\bot}^{2}}c_{0}-\left(  2b_{0}%
+c_{0}\right)  \lambda^{2}-2c_{0}\lambda^{2}\right]  z^{2}=0, \label{41}%
\end{gather}
which represents a coupling equation for the coefficients $\mathbf{A=}\left(
a_{0},b_{0},c_{0}\right)  ^{\text{T}}$
\begin{equation}
\mathbf{M}.\mathbf{A}=0, \label{42}%
\end{equation}
where the coupling matrix $\mathbf{M}$ is given by%
\begin{equation}
\mathbf{M=}\left(
\begin{array}
[c]{ccc}%
\frac{\omega^{2}}{\omega_{\bot}^{2}} & 2\left(  S^{2}-\lambda^{2}Z^{2}\right)
& \left(  S^{2}-\lambda^{2}Z^{2}\right) \\
0 & \left(  \frac{\omega^{2}}{\omega_{\bot}^{2}}-4\right)  & -1\\
0 & 2\lambda^{2} & \left(  \frac{\omega^{2}}{\omega_{\bot}^{2}}-3\lambda
^{2}\right)
\end{array}
\right)  , \label{M1}%
\end{equation}
Eq. (\ref{42}) has solutions as:%
\begin{equation}
\omega=0, \label{43}%
\end{equation}
which is corresponding to the constant shift of ground state density, and%
\begin{equation}
\omega^{2}\left(  m=0\right)  =\omega_{\bot}^{2}\left[  2+\frac{3}{2}%
\lambda^{2}\pm \frac{1}{2}\sqrt{9\lambda^{4}-16\lambda^{2}+16}\right]  .
\label{44}%
\end{equation}
The Eq. (\ref{44}) agree with the result in Ref.~(\cite{8}). As mentioned in
experiment Ref.~(\cite{24}), the frequency of collective excitation with
anisotropy $\lambda=\sqrt{8}$ is matched with theoretical results $\omega
_{-}\left(  \lambda=\sqrt{8}\right)  =1.797\omega_{\bot}$ very well. Eq.
(\ref{44}) can be reduced to the solutions for the quadrupole and monopole
excitations in the spherical trap when $\lambda \longrightarrow1$. For
cigar-type geometry ($\lambda \ll1$), the two frequencies become $\omega
_{+}=2\omega_{\bot}$ and $\omega_{-}=\sqrt{5/2}\omega_{z}$, which agree with
the experiment results in Ref.~(\cite{24}). While for disk-type geometry case
($\lambda \gg1$), the frequencies are: $\omega_{+}=\sqrt{3}\omega_{z}$ and
$\omega_{-}=\sqrt{10/3}\omega_{\bot}$.

Another interesting coupling phenomenon is that the surface mode
($n=0,\ell=3,m=0$) is coupled with radial-surface mode ($n=1,\ell=1,m=0$)
which will gives two decoupled mode, one of which is just dipole mode
($n=0,\ell=1,m=0$). As we know that dipole mode is corresponding to the center
of mass motion of BEC, thus, above coupling phenomenon means relative dynamics
of BEC may influence the center of mass motion and vice versa. By using the
same way as above, we can obtain the three decoupled modes as:%
\begin{equation}
\omega=\omega_{z}, \label{45}%
\end{equation}
which is corresponding to the one of dipole mode as mentioned before, and%
\begin{equation}
\omega^{2}\left(  m=0\right)  =\omega_{\bot}^{2}\left[  2+\frac{7}{2}%
\lambda^{2}\pm \frac{1}{2}\sqrt{25\lambda^{4}-16\lambda^{2}+16}\right]  ,
\label{46}%
\end{equation}
when $\lambda \longrightarrow1$, Eq.(\ref{46}) can be reduced to the isotropic
case $\sqrt{8}\omega_{0}$ and $\sqrt{3}\omega_{0}$, respectively. For
cigar-type geometry ($\lambda \ll1$), the frequencies become $\omega
_{+}=2\omega_{\bot}$ and $\omega_{-}=\left(  3\sqrt{2}/2\right)  \omega_{z}$.
While for disk-type geometry case ($\lambda \gg1$), the frequencies are:
$\omega_{+}=\sqrt{6}\omega_{z}$ and $\omega_{-}=\omega_{z}$ (which is equal to
dipole mode). Above mentioned "chain coupling dynamics" phenomenon may happen
more frequently when exciting the high energy collective mode.

\subsection{Non-Axial Symmetry}

In this section, we consider the case of the harmonic oscillator trap without
axial symmetry along the any axis: $V_{\text{ext}}=m\omega_{x}^{2}%
x^{2}/2+m\omega_{y}^{2}y^{2}/2+m\omega_{z}^{2}z^{2}/2$. In this case the
relevant hydrodynamic Eq. (\ref{11}) can be rewritten as:%
\begin{equation}
\omega^{2}\delta \rho=-\frac{1}{2}\nabla \cdot \left[  \omega_{x}^{2}\left(
x_{0}^{2}-x^{2}\right)  +\omega_{y}^{2}\left(  y_{0}^{2}-y^{2}\right)
+\omega_{z}^{2}\left(  z_{0}^{2}-z^{2}\right)  \nabla \delta \rho \right]  ,
\label{47}%
\end{equation}
where $m\omega_{x}^{2}x_{0}^{2}/2+m\omega_{y}^{2}y_{0}^{2}/2+m\omega_{z}%
^{2}z_{0}^{2}/2\equiv \mu$.

Since the trap has no axisymmetric, the magnetic quantum number $m$ of angular
momentum is no longer a good quantum number. Thus, the dispersion law will
depend on $m$ significantly. Although functions of the form Eq. (\ref{PolyEP})
are no longer the exact solutions of Eq. (\ref{47}) for $\delta \rho$,
excitation modes with different $m$ may be coupled together as results of
breaking rotational- and axial-symmetry in this case, however, for the
guidance with comparing with axial-symmetric case, we still find solution for
some specific value $\left(  n,l,m\right)  $. By using the same way as axial
symmetry case, the dispersion law we obtained as follows:%
\begin{equation}
\omega^{2}=\ell \bar{\omega}_{\perp}^{2}+E\left(  n,\ell \right)  \omega_{z}%
^{2}-F\left(  \ell,m\right)  \left(  \bar{\omega}_{\perp}^{2}-\omega_{z}%
^{2}\right)  +E\left(  n,\ell \right)  H\left(  \ell,m\right)  \left(
\bar{\omega}_{\perp}^{2}-\omega_{z}^{2}\right)  . \label{48}%
\end{equation}
where $\bar{\omega}_{\perp}^{2}\equiv \left(  \omega_{x}^{2}+\omega_{y}%
^{2}\right)  /2$, $E\left(  n,\ell \right)  \equiv \left(  2n^{2}+2n\ell
+3n\right)  $, $F\left(  \ell,m\right)  \equiv \left(  \ell+m\right)  \left(
\ell-m\right)  /\left(  2\ell-1\right)  $, and $H\left(  \ell,m\right)
\equiv \left(  \ell+m+2\right)  \left(  \ell+m+1\right)  /\left(  \left(
2\ell+1\right)  \left(  2\ell+3\right)  \right)  +\left(  \ell-m\right)
\left(  \ell-m-1\right)  /\left(  \left(  2\ell+1\right)  \left(
2\ell-1\right)  \right)  $ as defined in axial symmetry case. We notice that
here, Eq. (\ref{48}) has exact same form as Eq. (\ref{39}) in axial symmetry
case, except the transverse trapping frequency $\omega_{\bot}$ is now replaced
by the averaged transverse trapping frequency $\bar{\omega}_{\perp}$.

\subsection{Coupling Modes without Axial Symmetry}

Further, considering those excitation modes coupled together, the dispersion
law of the decoupled modes is obtained by the similar method as in axial
symmetry case. For example, if we consider following modes coupled together
as: ($n=1,\ell=0,m=0$) and ($n=0,\ell=2,m=\left(  0,\pm1,\pm2\right)  $), the
relevant hydrodynamic Eq. (\ref{11}) for $\delta \rho$ can be rewritten as:%
\begin{equation}
\delta \rho=aP_{\ell}^{2}+br^{2}Y_{2,0}+cr^{2}Y_{2,\pm2}+dr^{2}Y_{2,\pm1}%
=a_{0}+b_{0}x^{2}+c_{0}y^{2}+d_{0}z^{2}+e_{0}xz,\label{49}%
\end{equation}
by using the similar way as axial symmetry case, a coupling equation for the
coefficients $\mathbf{A=}\left(  a_{0},b_{0},c_{0},d_{0},e_{0}\right)
^{\text{T}}$
\[
\mathbf{M}.\mathbf{A}=0,
\]
here the coupling matrix $\mathbf{M}$ is given by:%
\begin{equation}
\mathbf{M=}\left(
\begin{array}
[c]{ccccc}%
\omega^{2} & 2\mu/m & 2\mu/m & 2\mu/m & 0\\
0 & \left(  \omega^{2}-3\omega_{x}^{2}\right)   & -\omega_{x}^{2} &
-\omega_{x}^{2} & f_{1}\\
0 & -\omega_{y}^{2} & \left(  \omega^{2}-3\omega_{y}^{2}\right)   &
-\omega_{y}^{2} & 0\\
0 & -\omega_{z}^{2} & -\omega_{z}^{2} & \left(  \omega^{2}-3\omega_{z}%
^{2}\right)   & f_{2}\\
0 & g_{1} & 0 & g_{2} & \left(  \omega^{2}-\omega_{x}^{2}-\omega_{z}%
^{2}\right)
\end{array}
\right)  .\label{50}%
\end{equation}
here, in order to illustrate the coupling caused by symmetry breaking, we have
introduced off-diagonal coupling term $\left(  f_{i}\text{, }g_{i}\right)  $
meaning y-component angular momentum has nonzero value. The description law
for this case is given by (without symmetry breaking terms):%
\[
\omega=0,
\]
which is corresponding to the constant shift of ground state density, and%
\begin{equation}
\omega^{2}=\omega_{x}^{2}+\omega_{z}^{2},\label{52}%
\end{equation}
which is corresponding to the collective excitation for degree $xz$, and Eq.
for the rest of degree
\begin{equation}
\omega^{6}-3\left(  \omega_{x}^{2}+\omega_{y}^{2}+\omega_{z}^{2}\right)
\omega^{4}+8\left(  \omega_{x}^{2}\omega_{y}^{2}+\omega_{x}^{2}\omega_{z}%
^{2}+\omega_{y}^{2}\omega_{z}^{2}\right)  \omega^{2}-20\omega_{x}^{2}%
\omega_{y}^{2}\omega_{z}^{2}=0,\label{53}%
\end{equation}
which has been mentioned in Ref.~(\cite{8})~as well.

As is shown in above, breaking symmetry may induce the coupling between modes
more seriously. For example, if considering the chiral symmetry broken due to
some external perturbation or spontaneously, following modes coupled concerned
as: ($n=1,\ell=0,m=0$) and ($n=0,\ell=2,m=\left(  0,+1,+2\right)  $), The
coupling matrix $\mathbf{M}$ is given by:%
\begin{equation}
\left(
\begin{array}
[c]{ccccccc}%
\omega^{2} & 2\mu/m & 2\mu/m & 2\mu/m & 0 & 0 & 0\\
0 & \left(  \omega^{2}-3\omega_{x}^{2}\right)   & -\omega_{x}^{2} &
-\omega_{x}^{2} & f_{1} & 0 & 0\\
0 & -\omega_{y}^{2} & \left(  \omega^{2}-3\omega_{y}^{2}\right)   &
-\omega_{y}^{2} & 0 & 0 & 0\\
0 & -\omega_{z}^{2} & -\omega_{z}^{2} & \left(  \omega^{2}-3\omega_{z}%
^{2}\right)   & f_{2} & 0 & 0\\
0 & g_{1} & 0 & g_{2} & \left(  \omega^{2}-\omega_{x}^{2}-\omega_{z}%
^{2}\right)   & 0 & 0\\
0 & 0 & 0 & 0 & 0 & \left(  \omega^{2}-\omega_{y}^{2}-\omega_{z}^{2}\right)
& h_{1}\\
0 & 0 & 0 & 0 & 0 & h_{2} & \left(  \omega^{2}-\omega_{x}^{2}-\omega_{y}%
^{2}\right)
\end{array}
\right)  \label{54}%
\end{equation}
here, in order to illustrate the coupling caused by nonzero value of
y-component angular momentum, we have introduced off-diagonal coupling term
$\left(  f_{i}\text{, }g_{i},h_{i}\right)  $, where, very interestingly,
additional term indicated by $h_{i}$ appearing due to the combined symmetry
broken. Here, dynamical degree of freedom is $\left(  1,x^{2},y^{2}%
,z^{2},xz,yz,xy\right)  $, which is corresponding to the coefficients
$\mathbf{A=}\left(  a_{0},b_{0},c_{0},d_{0},e_{0},f_{0},h_{0}\right)
^{\text{T}}$. The additional two frequency of excitation modes for the degree
$yz$ and $xy$ is given by (without symmetry breaking terms):%
\begin{equation}
\omega^{2}=\omega_{y}^{2}+\omega_{z}^{2},\label{55}%
\end{equation}
and
\begin{equation}
\omega^{2}=\omega_{x}^{2}+\omega_{y}^{2}.\label{56}%
\end{equation}
The same results can be obtained for the coupling between ($n=1,\ell=0,m=0$)
and ($n=0,\ell=2,m=\left(  0,-1,-2\right)  $).

\section{Anharmonic potential}

\label{F}

Through the developing the method and reviewing the collective excitation
provided in Sec. \ref{D} and \ref{E}, we then discuss the anharmonicity how to
influence the dispersion laws concerned. As mentioned before, in many
practical situation, the trapping potential is not accurately harmonic
\cite{30,30-1}, the condensed Bose gases are trapped with anharmonic potential
as: $V_{\text{ext}}\left(  r\right)  =m\omega_{0}^{2}\left(  r^{2}+\eta
r^{4}\right)  /2$, here, anharmonicity $\eta$ has dimension length$^{-2}$.
Through defining $\delta \rho=F\left(  q\right)  Y_{\ell m}$, where $q=r/R$,
the hydrodynamic Eq. (\ref{11}) can be rewritten as:
\begin{equation}
\omega^{2}\delta \rho=-\frac{1}{2}\omega_{0}^{2}\nabla \cdot \left[  R^{2}\left(
1-q^{2}-\tilde{\eta}q^{4}\right)  \cdot \nabla \delta \rho \right]  ,
\label{anharmonic}%
\end{equation}
here, $\tilde{\eta}=R^{2}\eta$ is a dimensionless parameter to characterize
effects of anharmonicity on the elementary excitation spectrum. By using the
same procedure as in Sec. \ref{D}, we can obtain the following equation
\begin{gather}
\frac{2\omega^{2}}{\omega_{0}^{2}}F\left(  q\right)  +\left(  \frac{\left(
1-q^{2}-\tilde{\eta}q^{4}\right)  }{q^{2}}\right)  \left[  \frac{\partial
}{\partial q}q^{2}\frac{\partial}{\partial q}F\left(  q\right)  -\ell \left(
\ell+1\right)  F\left(  q\right)  \right] \nonumber \\
-\left(  2q+4\tilde{\eta}q^{3}\right)  \frac{\partial}{\partial q}F\left(
q\right)  =0. \label{58}%
\end{gather}
As mentioned before, anharmonicity $\tilde{\eta}$ can significantly cause
center-of-mass and relative motion coupled, resulting in effective changing
the collective excitation spectrum. Seeking the same form of solution as that
in harmonic case (\ref{PolyEP}), based on Eq. (\ref{58}), we can find a
coupled equation as:%
\begin{gather}
\alpha_{2n}\left[  \frac{2\omega^{2}}{\omega_{0}^{2}}-2n\left(  2n+3\right)
-\ell \left(  \ell+1\right)  \right]  +\alpha_{2n-2}\left[  -\left(
2n-2\right)  \left(  2n+3\right)  +\ell \left(  \ell+1\right)  \right]
\tilde{\eta}=0,\nonumber \\
\alpha_{2n}\left[  2n\left(  2n+1\right)  -\ell \left(  \ell+1\right)  \right]
+\alpha_{2n-2}\left[  \frac{2\omega^{2}}{\omega_{0}^{2}}-\left(  2n-2\right)
\left(  2n+1\right)  -\ell \left(  \ell+1\right)  \right]  =0.
\label{anharmonicCE}%
\end{gather}

\subsection{Surface modes}

First, we consider the lowest radial modes $\left(  n=0\right)  $, which is
also called surface excitations. Starting from Eq. (\ref{anharmonicCE}), we
predict the dispersion law as:%
\begin{equation}
\omega_{s,\pm}\left(  n=0\right)  =\frac{1}{\sqrt{2}}\sqrt{-1+\ell+\ell^{2}%
\pm \sqrt{1-6\ell \tilde{\eta}-7\ell^{2}\tilde{\eta}-2\ell^{3}\tilde{\eta}%
-\ell^{4}\tilde{\eta}}}\omega_{0}, \label{anharDP1}%
\end{equation}
here, surface excitation spectrum is now separated into two branches, which is
quite different from the harmonic trapping case. In particular, anharmonicity
$\tilde{\eta}$ has a critical value%
\begin{equation}
\tilde{\eta}_{s,c}=\frac{1}{\ell \left(  6+7\ell+2\ell^{2}+\ell^{3}\right)  },
\label{anharSM1}%
\end{equation}
above which surface excitation would unstable which means anharmonic induced
instability of surface excitation. Moreover, $\tilde{\eta}_{s,c}$ is always
positive value for any multipole mode. Importantly, for higher multipole mode
$\ell$, the critical value of $\tilde{\eta}_{s,c}$ decreases very rapidly (for
example, $\tilde{\eta}_{s,c}\approx0.0046$ for octupole mode $\ell=3$), which
means a little bit of anharmonic perturbation will cause the higher multipole
mode damped or blowed up very rapidly, depending on which branches mode
belongs to. Another interesting phenomena is, although the frequency of plus
branch of these modes $\left(  \omega_{s,+}\right)  $ is always higher than
the value $\sqrt{\ell}\omega_{0}$ of harmonic trapping case, the frequency of
minus branch of these modes $\left(  \omega_{s,-}\right)  $ can be lie
systematically below or above the harmonic trapping result, depending on the
value of anharmonicity $\tilde{\eta}$ and of angular momentum $\ell$ of the
excitation, where critical value is given by%
\begin{equation}
\tilde{\eta}_{s,s}=\frac{-2+3\ell-\ell^{2}}{\left(  6+l+\ell^{2}\right)  },
\label{anharSM2}%
\end{equation}
here, $\tilde{\eta}_{s,s}$ is always negative except dipole and quardupole
case $\left(  l=1,2\right)  $, above which the frequency of minus branch of
these modes is always lower than those of harmonic trapping case (for example,
for octupole mode, $\tilde{\eta}_{s,s}=-1/9$, $\omega_{s,-}\left(
n=0,\ell=3,\tilde{\eta}=-2/9\right)  =\sqrt{2}\omega_{0}$). When considering
the influence of instability critical value $\tilde{\eta}_{s,c}$ and lower
branch critical value $\tilde{\eta}_{s,s}$ jointly, in order to existing the
stable surface mode with lower frequency than the harmonic trapping case, the
condition for the anharmonicity $\tilde{\eta}$ should be satisfied as
following
\begin{equation}
\tilde{\eta}<\tilde{\eta}_{s,c}\text{ and }\left \vert \tilde{\eta}%
_{s,s}\right \vert <\left \vert \tilde{\eta}\right \vert , \label{anharSM3}%
\end{equation}
which holds for any given angular momentum $\ell$.

\subsection{Compression modes}

In a similar way, starting from Eq. (\ref{anharmonicCE}), we can determine the
frequency of compression modes which is also separated into two branches as:%
\begin{equation}
\omega_{c,\pm}\left(  l=0\right)  =\frac{1}{\sqrt{2}}\sqrt{-1+2n+4n^{2}%
\pm \sqrt{1-16n^{3}\tilde{\eta}-16n^{4}\tilde{\eta}+4n(2+3\tilde{\eta}%
)+4n^{2}(4+5\tilde{\eta})}}\omega_{0}, \label{anharDP2}%
\end{equation}
here, very interestingly, our monopole mode $\omega_{c,\pm}\left(
n=1,l=0\right)  $ is the same as harmonic trapping case ($\sqrt{5}\omega_{0}$)
for any anharmonic perturbation $\tilde{\eta}$. In particular, similar to the
surface mode case, anharmonicity $\tilde{\eta}$ has critical value%
\begin{equation}
\tilde{\eta}_{c,c}=\frac{(4n+1)^{2}}{4n(-3-5n+4n^{2}+4n^{3})},
\label{anharCM1}%
\end{equation}
above which compression excitation would unstable, which also reflects
anharmonic induced instability of compression excitation. Similarly,
$\tilde{\eta}_{c,c}$ is always positive for any compression mode except
monopole mode $n=1$. Especially, for higher mode $n$, the critical value of
$\tilde{\eta}_{c,c}$ also decreases very faster with increasing compression
mode $n$, (for example, $\tilde{\eta}_{c,c}\approx0.117$ for corresponding
mode $n=3$), which means a little bit of anharmonic perturbation will cause
the higher multipole mode damped or blowed up very rapidly, depending on which
branches mode belongs to. However, very different from the surface mode case,
the frequency of plus and minus branch of anharmonic compression modes
$\left(  \omega_{c,\pm}\right)  $ can be both significantly lower than the
value $\sqrt{2n^{2}+3n}\omega_{0}$ of harmonic trapping case, where critical
anharmonic value for plus and minus branch have the same value which is just%
\begin{equation}
\tilde{\eta}_{c,s}=0, \label{anharC2}%
\end{equation}
above which the frequency of compression modes is lower than those of harmonic
trapping case, for example, for radial mode $n=2$, we can obtain $\tilde{\eta
}_{c,c}=0.289$, and plus and minus branch are $\omega_{c,+}\left(
n=2,\ell=0,\tilde{\eta}=0.2\right)  =3.464\omega_{0}$, and $\omega
_{c,-}\left(  n=2,\ell=0,\tilde{\eta}=0.2\right)  =2.646\omega_{0}$,
respectively, if anharmonicity $\tilde{\eta}=0.2<\tilde{\eta}_{c,c}$, but if
tuning the anharmonicity $\tilde{\eta}=0.3>\tilde{\eta}_{c,c}$, plus and minus
branch would unstable which are conjugated each other and $\omega_{c,+}\left(
n=2,\ell=0,\tilde{\eta}=0.3\right)  =\omega_{c,-}^{\ast}=(3.085-0.140i)$,
similar to the damped or blowed up mode in surface excitation. When
considering the influence of instability critical value $\tilde{\eta}_{c,c}$
and two branch critical value $\tilde{\eta}_{c,s}$ jointly, in order to
existing the stable compression mode with lower frequency than the harmonic
trapping case, the condition for the anharmonicity $\tilde{\eta}$ should be
satisfied as following
\[
\tilde{\eta}_{c,s}\text{ }\left(  =0\right)  <\tilde{\eta}<\tilde{\eta}%
_{c,c},
\]
which holds for any given quantum radial number $n>1$.

\section{Conclusion}

\label{G}

In summary, we studied anharmonicity-induced critical behaviors on the
collective excitation spectrum of 3D BEC. We found the dispersion law is now
separated into two branches, and appear two limiting cases depending on the
anharmonicity $\tilde{\eta}$ of the BEC. For a upper limit, BEC can be
unstable with respect to some specific collective excitation, while for the
lower limit, the anharmonicity-influenced frequency of collective excitation
can be effectively lower than that without anharmonicity. Our results
demonstrate that the effects of anharmonicity can play key role on exciting
elementary excitations which can be readily reached upon the recently
experimental tools and techniques \cite{30,30-1}. Extending the radial
anharmonicity to angular anharmonicity of trapping potential could be
investigated in the future.

The financial support from the early development program of NanChang University and Skoltech-MIT Next
Generation Program is gratefully acknowledged.

\end{document}